\documentclass[]{beilstein}

\usepackage{xspace}

\usepackage{amssymb,amsbsy,amsmath}

\newcommand{\fmref}[1]{(\protect\ref{#1})}

\begin{document}

\title{Classical molecular dynamics investigations of
  biphenyl-based carbon nano membranes}
\author{Andreas Mrugalla}
\author*{J\"urgen Schnack}{jschnack@uni-bielefeld.de}
\affiliation{Fakult\"at f\"ur Physik, Universit\"at Bielefeld, Postfach 100131, D-33501 Bielefeld, Germany}
\maketitle

\begin{abstract}
\background Free-standing carbon nanomembranes (CNM) with molecular
thickness and macroscopic size are fascinating objects both for
fundamental reasons and for applications in
nanotechnology. Although being made from simple and identical
precursors their internal structure is not fully known and
hard to simulate due to the large system size that is necessary
to draw definite conclusions.
\results We performed large-scale classical molecular dynamics
  investigations of biphenyl-based carbon nanomembranes. We
  show that one-dimensional graphene-like stripes constitute a
  highly symmetric quasi one-dimensional ground state. This
  state does not crosslink. Instead crosslinked structures are
  formed from highly excited precursors with a sufficient amount
  of broken phenyls. 
\conclusion The internal structure of CNM is very likely a
disordered metastable state which is formed in the process of
cooling. 
\end{abstract}

\keywords{biphenyls; carbon nanomembranes; classical molecular dynamics}

\section{Introduction}

Freestanding carbon nanomembranes are produced from molecular
precursors such as for instance biphenylthiols. They self-assemble in
monolayers on gold surfaces if polymerized by radiation with
electrons \cite{GSE:APL99,TBN:AM09,AVW:ASCN13}. The product is a membrane,
whose thickness, homogeneity and surface chemistry are 
directly related to the molecular precursor. So far several
classes of precursors have been exploited \cite{AVW:ASCN13}.

One of the major unsolved questions is the internal structure of
these membranes, since the structure cannot be determined by
x-ray. In Ref.~\cite{TKE:L09} quantum chemical calculations were
performed for various dimers of biphenyls, 
which left open how the precursor molecules interlink
laterally. A first small-scale quantum calculation (using ARGUS Lab)
of a two-dimensional cutout of 6 by 5 
biphenyls is reported in Ref.~\cite{TKE:L09}. These calculations
suggest that the regular 
structure of the precursor SAM turns into a disordered
sheet. Nevertheless, the simulations of the very small system
do not allow definite conclusions about the structure of the extended
sheet. On the other hand, the quantum mechanical simulation of
extended systems even by means of Density Functional Theory (DFT)
has to assume a regular lattice and can treat only small unit
cells \cite{HRZ:NL07,HRB:L08,CAS:PCCP10}. Consequently, the
resulting structure is also regular \cite{CAS:PCCP10}. If one,
as in the present case, 
can expect that the structure is irregular, i.e. a lattice
structure as in solids cannot be assumed, a quantum mechanical
simulation is virtually impossible. 

In this article we therefore resort to Classical Molecular
Dynamics (CMD) simulations which allow to simulate up to several
millions of carbon atoms. In order to account for the very
flexible $sp^n$--binding modes of carbon we use the modern carbon-carbon
potential of Nigel Marks \cite{Mar:PRB00} which has been
demonstrated to be able to simulate extended carbon structures
\cite{MCM:PRB02,PML:PRB09}. We focus our investigations on CNMs
made of biphenylthiols. The simulational results of our energy
minimizing procedure yield -- depending on the initial state --
a large variety of structures, among which a very regular one
made of parallel graphene stripes has the lowest energy. Our
hypothesis is, that in a
realistic synthesis process such an idealized state is not
reached, instead the system ``freezes" into a metastable
irregular configuration that is laterally linked through carbon bonds
of broken phenyls. We show that such structures indeed form in
our simulations.

The article is organized as follows. In the next section we
shortly repeat the essentials of our classical molecular
dynamics simulations. The main section discusses the
results. The article closes with an outlook.

\section{Classical carbon-carbon interaction}
\label{sec-2}

\begin{figure}
\centering
\includegraphics*[clip,width=60mm,keepaspectratio]{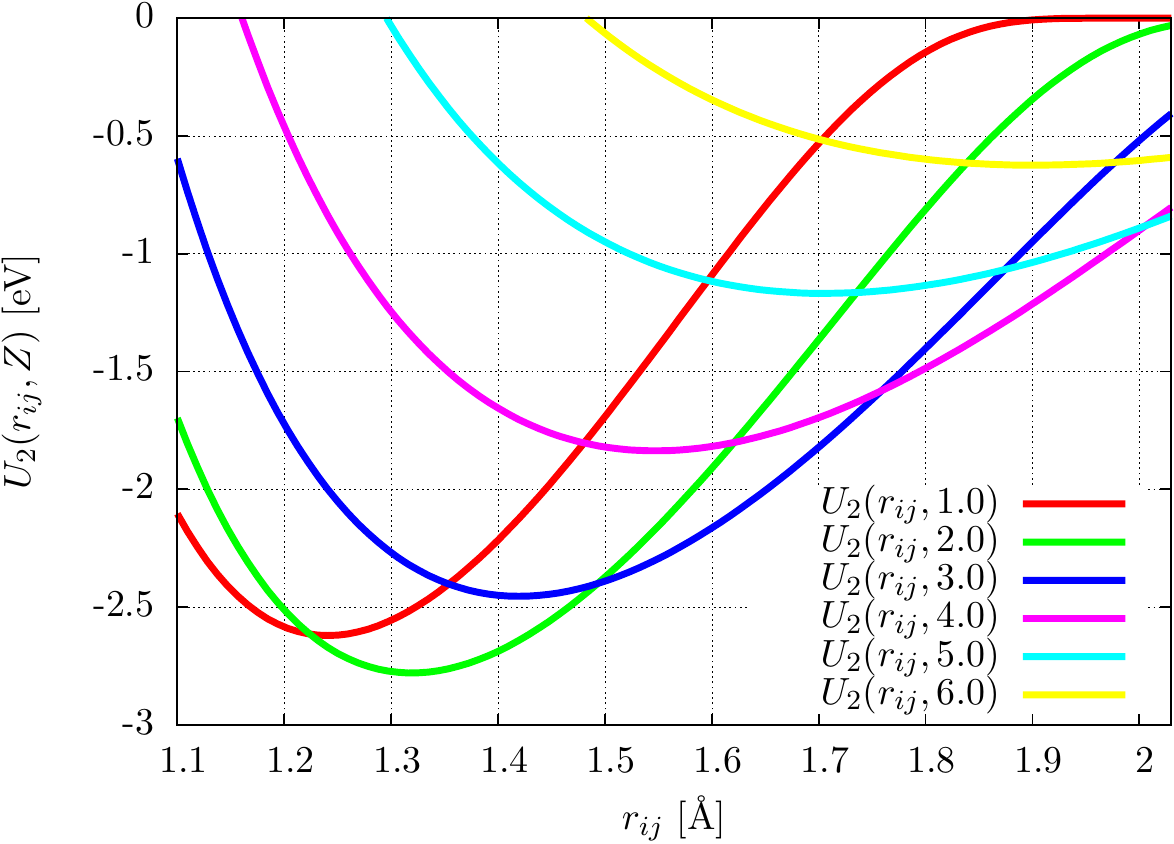}
\quad
\includegraphics*[clip,width=60mm,keepaspectratio]{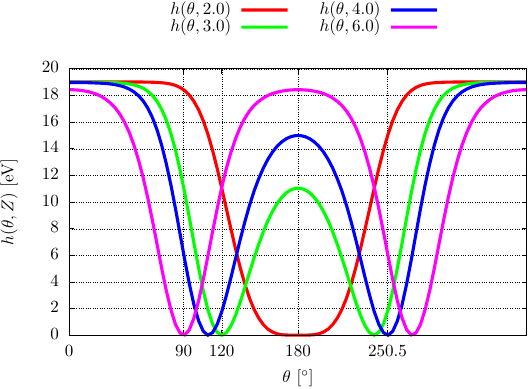}
\caption{Pictorial representation of the major ingredients of
  the carbon-carbon potential \fmref{eq:U}, compare also
  Ref.~\cite{Mar:PRB00}. L.h.s.: two-body 
  potential for various coordination numbers. R.h.s.:
  coordination number dependence of angular-dependent part of
  the three-body potential.} 
\label{classical-cnm-a}
\end{figure}

A realistic classical carbon-carbon interaction must be able to
account for the various $sp^n$--binding modes. Two 
potentials, developed by Tersoff and Brenner, have been used for
carbon materials as well as for hydro-carbons
\cite{Ter:PRB88,Bre:PRB90,BSH:JPCM02}. In our investigations we
employ the improved potential by Marks
\cite{Mar:PRB00}. This potential comprises density-dependent
two- and three-body potentials, $U_2$ and $U_3$ respectively, 
\begin{equation}
V\left(\vec{R}_1,\dots,\vec{R}_N\right) = \sum_{i=1}^{N} \left(
\sum_{\substack{j=1\\j\neq i}}^{N} U_2(R_{ij}, Z(i)) +
\sum_{\substack{j=1\\j\neq i}}^{N} \sum_{\substack{k=j+1\\k\neq
    i}}^{N} U_3(R_{ij},R_{ik},\theta(i,j,k),
Z(i))\right)\label{eq:U} 
\end{equation}
which account for the various binding modes. We would not like
to repeat the technical details, which are given in
Ref.~\cite{Mar:PRB00}, but rather show with two figures how such
effective potentials work. Figure~\ref{classical-cnm-a} shows on
the l.h.s. the radial dependence of the two-body potential for
various coordinations $Z(i)$, i.e. various numbers of nearest
neighbor atoms. The general trend is that the 
bond weakens and the minimum shifts to larger distances with
coordination. On the r.h.s. of Fig.~\ref{classical-cnm-a} a
major ingredient to the three-body term is shown which regulates
the bonding angles. As one can see, a single carbon with two
neighbors leads to a linear configuration, with three neighbors
a $120^\circ$-configuration is assumed, and so on.

We tested the potential for several typical $sp^2$--bonded
materials such as graphene and carbon nanotubes and obtained 
perfect structures \cite{Mrugalla:Master13}. It should be made
clear at this point that classical molecular dynamics cannot
describe electronic properties or molecular orbitals, but
structure in the sense of atomic positions and mechanical
properties such as vibrational spectra or Young's moduli.

\section{Results and Discussion}
\label{sec-3}

Structure investigations have been performed for arrangements of
$10 \times 10$ biphenyls, i.e. 1200 carbon atoms. The initial
state was assumed either regular in various configurations
including twists between lower and upper phenyls or random with
small fluctuations around original carbon positions. An energy
minimization was performed in order to reach a local energy
minimum. That the realistic CNM state is described by a local
energy minimum is an important conceptual ingredient of the modeling. The
global energy minimum would be given by the graphene structure,
which is not reached in the course of the synthesis process, but
can indeed be reached experimentally by heating the material
\cite{TKE:L09}. 

\begin{figure}
\centering
\includegraphics*[clip,width=100mm,keepaspectratio]{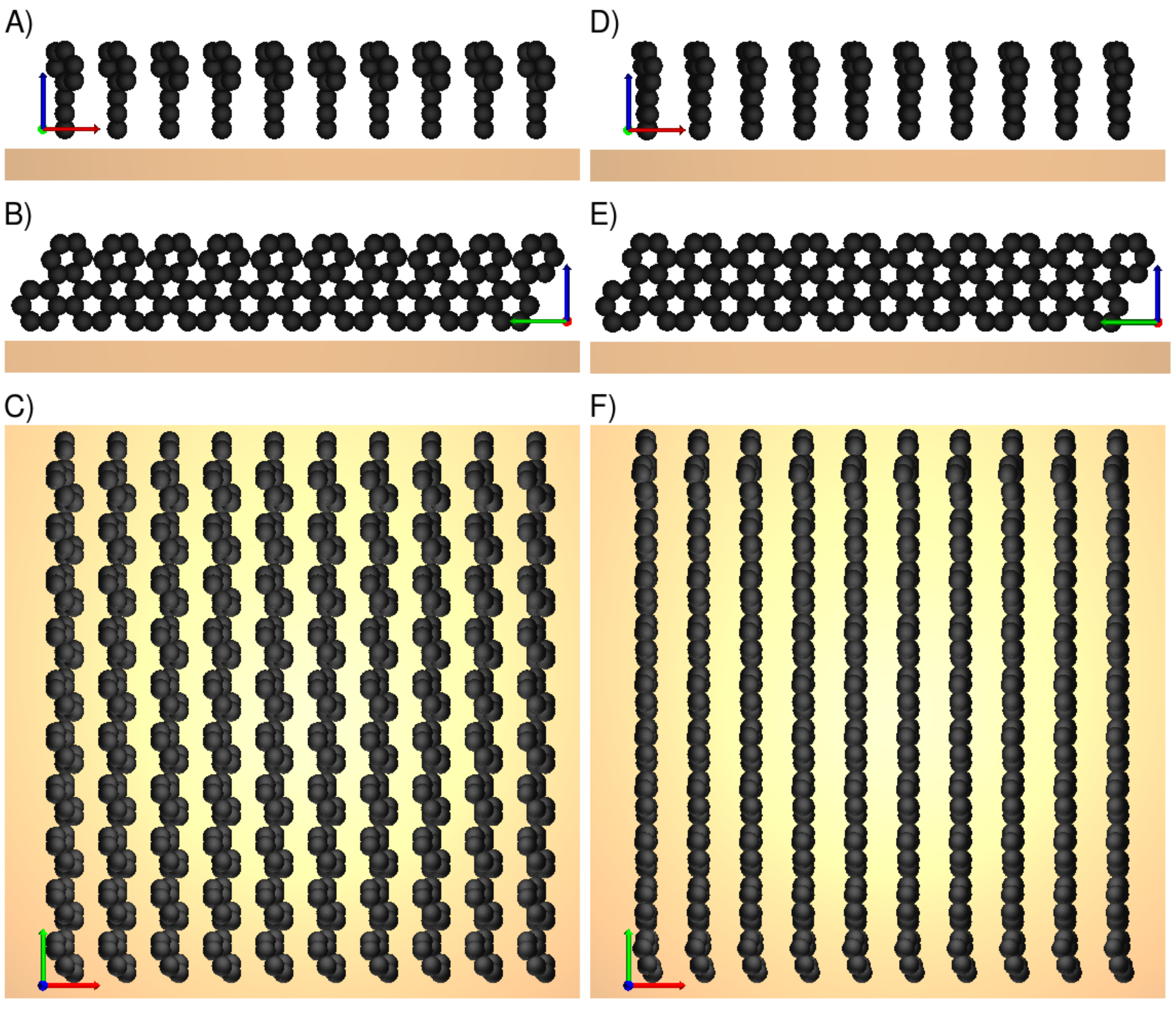}
\caption{Front (A), side (B) and top (C) view of an initial
  state made of a regular arrangement of tilted 
  biphenyls. The tilt angle is $30^\circ$. Front (D), side (E)
  and top (F) view of the resulting local energy minimum state,
  which consists of regular graphene stripes.} 
\label{classical-cnm-b}
\end{figure}

It turns out that a graphene-like structure constitutes a deep
local energy minimum. This structure is given by parallel
graphene stripes that form for a tilt angle of $30^\circ$ with
respect to the surface normal, compare
Fig.~\ref{classical-cnm-b}. The stripe structure forms even if the
initial state is moderately excited by carbon displacements
about their mean positions. It is interesting to note that a
similar angle of $31^\circ$ was determined by means of Near-Edge
X-ray Absorption Spectroscopy (NEXAFS) investigations on
pristine monolayers \cite{GSE:APL99,FRH:L02,TKE:L09}. This angle
seems to increase, on average, to $41^\circ$ after irradiation
\cite{TKE:L09}. 

\begin{figure}
\centering
\includegraphics*[clip,width=100mm,keepaspectratio]{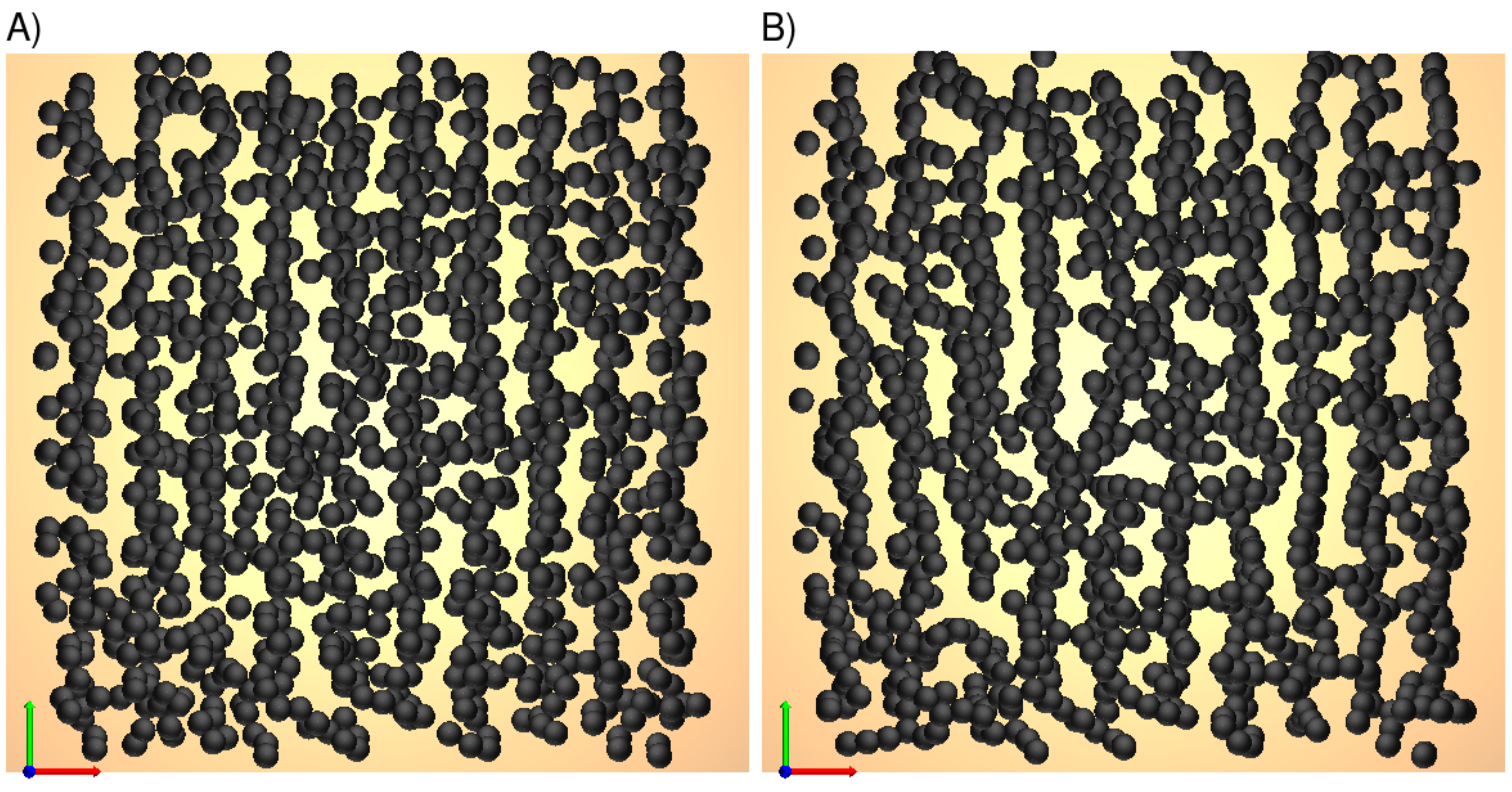}
\caption{Top view of a strongly excited initial state (A) and of
  the corresponding local energy minimum state (B).} 
\label{classical-cnm-c}
\end{figure}

We conjecture that the also observed amount of 
destroyed phenyl bonds, Ref.~\cite{TKE:L09}, plays an important
role in understanding the formation of laterally interlinked
biphenyls. It signals that the CNM is very likely laterally
interlinked through broken
phenyls. Figure~\fmref{classical-cnm-c} shows on the l.h.s.
as an example the initial randomized configuration as it could
be realized during the electron exposition and on the r.h.s. the
resulting state found by steepest descent.

We performed about 50 of the very demanding simulations for various
initial conditions characterized by the structure of the
two-dimensional lattice, by the tilt angle, the torsion angles
of the phenyls, and by the initial
displacements of the carbon atoms from their biphenyl
positions. Although these degrees of freedom constitute an
enormous parameter space, we are able to summarize our numerical
experience like follows: For local
displacements of less than about $\pm 1$~\AA\ always the graphene stripe
configuration forms. This is in part also observed in DFT
calculations \cite{CAS:PCCP10}. Only for sufficient randomization of
the initial state, which corresponds to a substantial excitation and to the
breakup of sufficiently many phenyls, a true
crosslinking is observed. The resulting states are characterized
by an irregular structure with pores of various sizes, as can be
seen in Fig.~\ref{classical-cnm-c} (B).

\section{Outlook}
\label{sec-4}

Our investigations demonstrate that carbon nanomembranes, which
are produced from molecular precursors such as for instance
biphenylthiols, very likely constitute irregular metastable
configurations that form from highly excited randomized
self-assembled monolayers. This suggests that the electron
exposition (dose, time, energy) as well as the cooling
dynamics plays an important role for the actual
structure. Future investigations will focus on these aspects as
well as on obvious consequences such as defect formation,
e.g. pores.

\begin{acknowledgements}
We are very thankful to Prof. Nigel Marks for sharing with us
the details of his carbon-carbon potential and Daniela Ramermann
by supporting us with povray figures. 
\end{acknowledgements}


\vspace{3cm}
This article is published in full length in \textit{Beilstein
  J. Nanotechnol.}
\textbf{20??}, \textit{?}, No. ?.

\end{document}